\documentclass{iau}
\usepackage{graphicx}

\usepackage{natbib,amsmath,amssymb,hyperref}
\usepackage{comment}

\newcommand{\orcidicon}[1]{\href{https://orcid.org/#1}{\includegraphics[width=11pt]{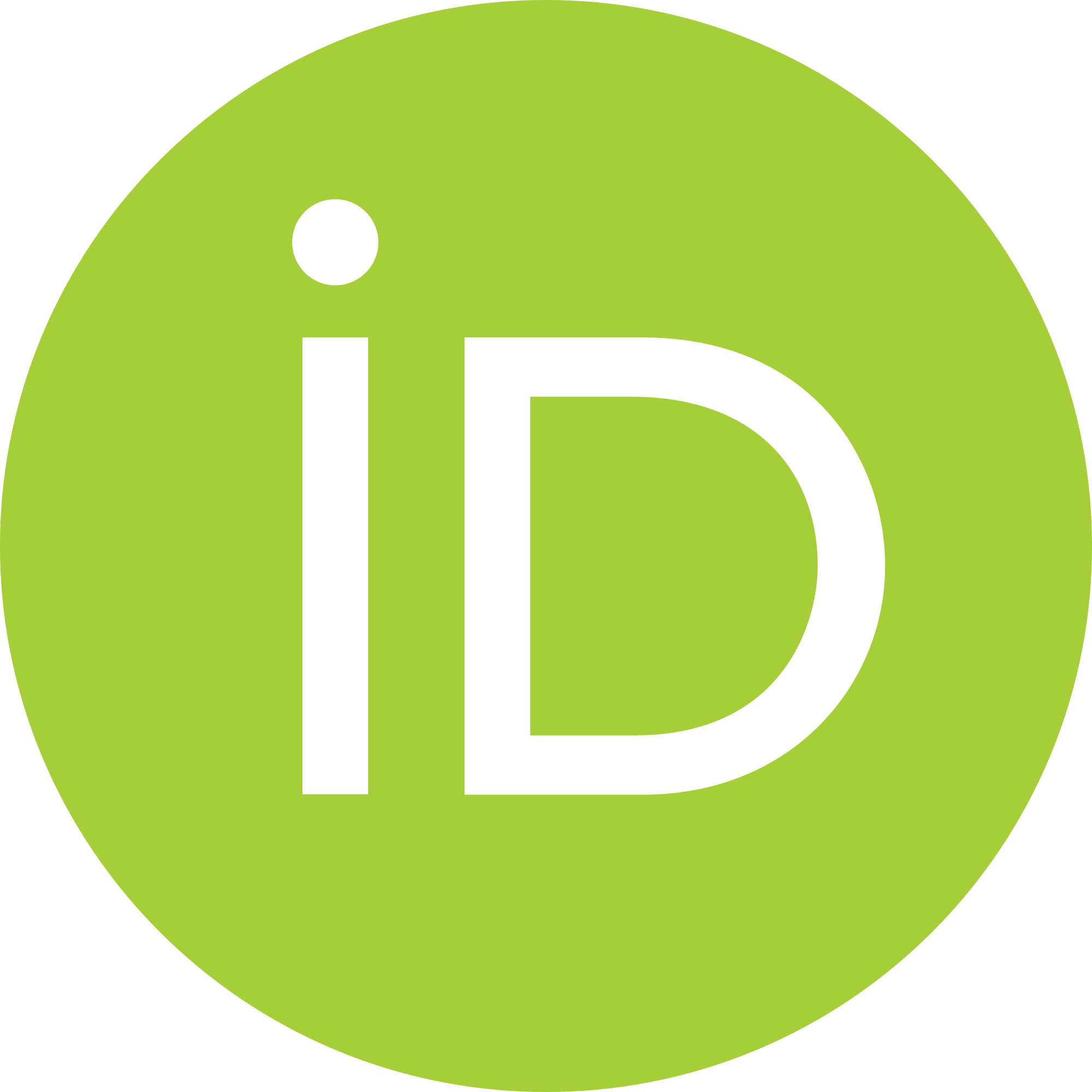}}}
\newcommand{\orcid}[1]{\href{https://orcid.org/#1}{\protect\orcidicon{#1}}}

\title[Hierarchical BBH Mergers in AGN Disks] 
{Hierarchical Black Hole Mergers in AGN
Disks: Tracing Massive Black Hole Growth Across
Cosmic Time}

\author[M. Paola Vaccaro]   
{Maria Paola Vaccaro \orcid{0000-0003-3776-9246}
 }

\affiliation{Universit\"at Heidelberg, \\ Zentrum f\"ur Astronomie (ZAH), Institut f\"ur Theoretische Astrophysik, \\ Albert-Ueberle-Str. 2, 69120, Heidelberg, Germany \\ email: {\tt mariapaolavaccaro@gmail.com} }

\pubyear{2025}
\volume{398}  
\pagerange{??}
\jname{MODEST-25}
\editors{Hyung Mok Lee, Rainer Spurzem and Jongsuk Hong}
\begin{document}

\maketitle

\begin{abstract}
The dense and gaseous environments of active galactic nuclei (AGNs) can catalyze repeated mergers of stellar-mass black holes (BHs), potentially explaining the high-mass tail of binary black hole (BBH) mergers observed by LIGO-Virgo-KAGRA (LVK). We present a semi-analytical population synthesis framework that captures key physical processes in AGN disks, including gas capture, migration, binary pair-up, gas hardening, and dynamical binary-single interactions. Our simulations show that AGN disks can produce hierarchical mergers, especially near migration traps, and may contribute to the high-mass, high-spin BBH population. This work opens prospects for constraining both the AGN channel and supermassive black hole growth with future gravitational-wave detections.
\keywords{black hole physics, stars: black holes, stars: kinematics and dynamics, gravitational waves, galaxies: active}
\end{abstract}


The LVK collaboration has detected merging BBHs with masses extending into the pair-instability mass gap ($\gtrsim 60\,M_\odot$), suggesting a dynamical origin \citep{Abbott2023}. One promising environment for such mergers is the accretion disk surrounding a supermassive black hole (SMBH) in an AGN. Gas-assisted dynamics favor a high encounter rate between BHs, enhancing the likelihood of binary formation. Moreover, remnants of previous mergers are retained thanks to the deep gravitational potential well of the SMBH, leading to multiple generations of mergers 
\citep[e.g.][]{Tagawa2020}.


We present a fast, flexible framework for simulating BBH mergers in AGN disks that reproduces key features of hierarchical mergers and can probe the origin of extreme-mass, high-spin BBHs. Moreover, the redshift evolution of AGN-driven merger rates can shed light on SMBH growth across cosmic time.

\section{Semi-Analytical Model for BBH Mergers in AGN Disks}

We model BBH formation and mergers in AGN disks by integrating the relevant physical processes into a consistent population-synthesis framework, based on the \textsc{fastcluster} code \citep{Mapelli2021} extended with AGN-specific prescriptions \citep{Vaccaro2024}. The AGN disk is described using the steady-state model of \citet{Sirko2003}, which provides radial profiles of gas surface density $\Sigma_\mathrm{g}$ and aspect ratio $h$. These profiles are computed numerically with the \textsc{pAGN} code \citep{Gangardt2024} for given values of the SMBH mass $M_\bullet$ and the disk viscosity parameter $\alpha$.
We consider stellar-origin BHs formed in a coexisting nuclear star cluster, initialized at solar metallicity using the \textsc{SEvN} code \citep{Iorio2023}. BHs are distributed uniformly in radius with isotropic initial inclinations. We follow the evolution of BHs individually, computing key timescales analytically.
Such BHs are captured by the disk via gas drag 
\citep[timescale computed as in][]{Wang2024}. Once embedded, they undergo 
migration, which transports them towards locations of higher BH numerical density and favors binary formation \citep{Bellovary2016}. We model this analytically 
following \citet{McKernan2012}. If the resulting timescale is shorter than the AGN lifetime $\tau$, we consider BBH formation and extract the mass $m_2$ of the secondary BHs by physically-motivated distributions \citep{Vaccaro2024}. 

We show how the process of binary formation is sensitive to the component masses and the local disk properties. Migration traps, defined as locations where torques cancel and migration transitions from outwards to inwards, are favorable locations for BBH pair-up, but other locations in the disk are also viable sites for BBH assembly. The proportion between trap and non-trap mergers depends on the SMBH mass $M_\bullet$ and the gas viscosity $\alpha$ \citep{Vaccaro2025}.

After formation, BBHs evolve under three main processes:  
(i) \emph{Gas hardening}, which exchanges angular momentum with the surrounding gas and is modeled following \citet{Ishibashi2024}. In thin-disk regions ($h \lesssim 0.04$--$0.16$ depending on the binary accretion rate) this shrinks the binary and excites its eccentricity; in thicker regions, it tends to widen it.  
(ii) \emph{Three-body encounters} with single BHs, modeled by interfacing our code with \textsc{tsunami} \citep{Trani2024}, can lead to exchanges, orbital tilts, and changes in semi-major axis and eccentricity \citep{Samsing2022, Trani2023}. These interactions generally shrink BBHs and increase eccentricity. 
(iii) \emph{Gravitational-wave emission}, computed via the \citet{Peters1964} formalism, drives the final inspiral and merger.
  
The remnant mass $m_\mathrm{rem}$ and spin $\chi_\mathrm{rem}$ are computed from numerical-relativity fits \citep{Jimenez2017}. Remnants may undergo further capture, migration, and binary formation, enabling hierarchical mergers, if three conditions are met: the recoil velocity $v_\mathrm{kick}$ is below the local escape velocity $v_\mathrm{esc}$ (met in $99.4\%$ of cases); the cumulative accreted mass $M_\mathrm{acc}=m_1+\sum m_2^{(N)}$ does not exceed the total captured BH mass in the disk ($72.9\%$); and the remnant’s formation timescale is shorter than $\tau$ ($41.3\%$).  
When satisfied, this produces a chain of progressively more massive BBH mergers \citep{Vaccaro2024}.

\begin{figure}
    \centering
    \includegraphics[width=\linewidth]{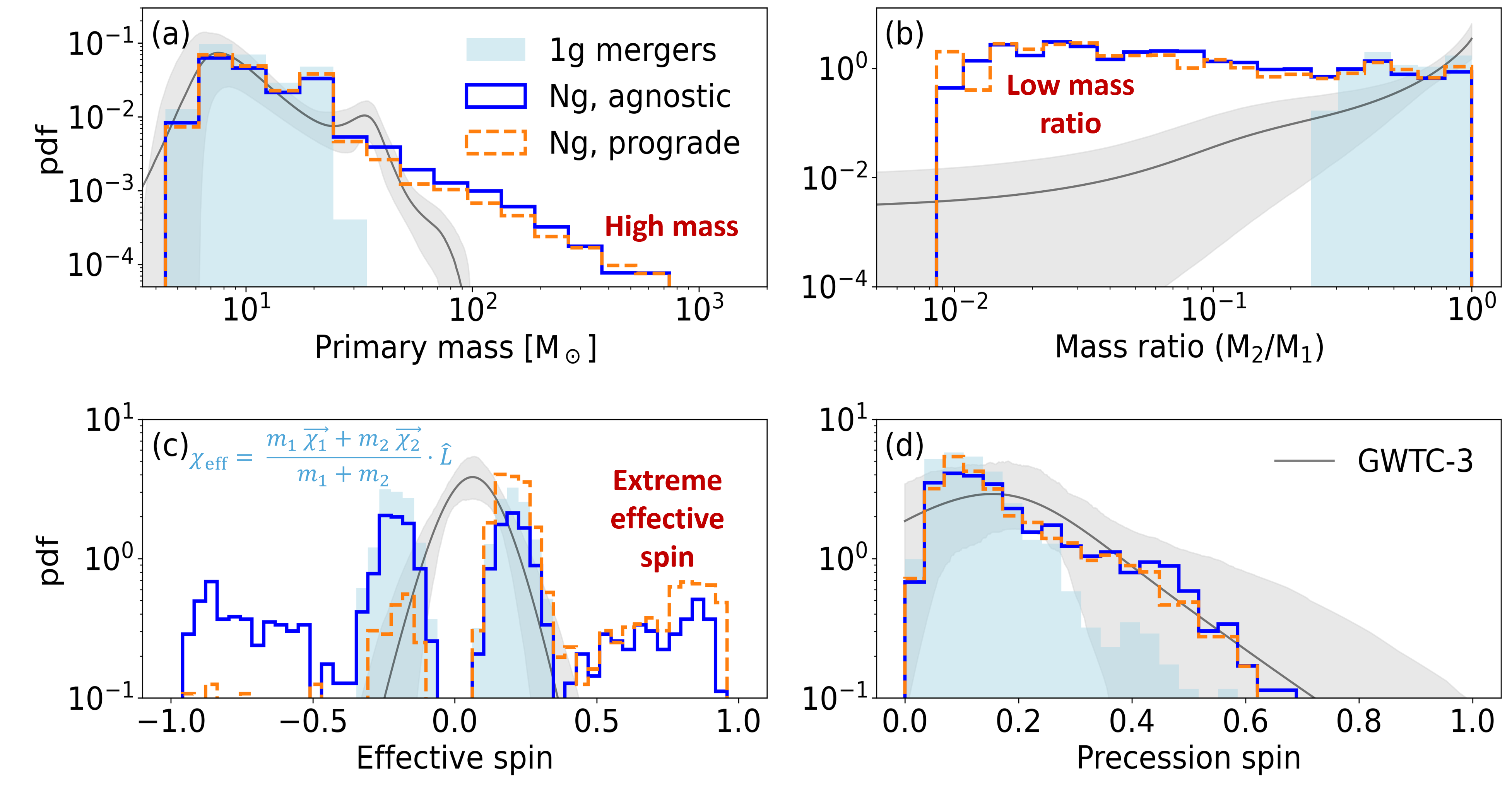}
    \caption{
Distributions of BBH merger properties from our AGN channel models. 
(a) Primary mass $m_1$, (b) mass ratio $q=m_2/m_1$, (c) effective spin $\chi_\mathrm{eff}$, and (d) precession spin $\chi_\mathrm{p}$. 
Solid blue and dashed orange lines show hierarchical mergers including all orbital orientations (``agnostic'') and restricted to prograde orbits at BBH formation (``prograde''), respectively. 
Light blue histograms show 1g mergers from our simulations. 
Grey curves and shaded regions indicate the median and $90\%$ credible intervals from GWTC-3 \citep{Abbott2023}. 
}
    \label{fig:slide11}
\end{figure}

\section{Results and Predictions}

\subsection{Population of hierarchical mergers in AGNs}
As shown in \autoref{fig:slide11}, our simulations show that AGN disks can efficiently produce a distinct population of hierarchical BBH mergers. This population extends the primary mass distribution well beyond $100M_\odot$, with a significant fraction reaching into and above the pair-instability mass gap \citep{Vaccaro2024}. Mass ratios are often low ($q \lesssim 0.1$), reflecting the frequent pairing of massive merger remnants with lighter companions, and effective spins span extreme values ($\chi_{\rm eff} \simeq \pm 0.8$) due to repeated mergers and dynamical re-orientations. Spin and orbital properties are further diversified by three-body interactions, which can tilt spins, increase eccentricities, and in some cases exchange binary members.

\subsection{Constraining the origin of observed BBH mergers}
We assess the possible contribution of the AGN channel to the current GW detections by performing a hierarchical Bayesian mixture analysis including multiple formation channels. 
We use \textsc{fastcluster} predictions and evaluate transient event detectability for the LVK network as in \citet{Princess}. The mixing fraction $f_\mathrm{AGN}$ is then inferred by comparing the model distributions to the posterior samples of BBH events in \citet{Abbott2023}. The inferred $f_\mathrm{AGN}$ in our analysis has a median value of $\sim 8\%$ with a $90\%$ credible interval spanning from $<1\%$ up to $\sim 20\%$ \citep{Vaccaro2024}. It is lower than for other dynamical channels (e.g. star clusters), but remains consistent with a non-negligible contribution of AGNs to the observed BBH population. A comparable upper limit ($f_\mathrm{AGN} \lesssim 17\%$ for bright AGNs) has been independently obtained via a spatial correlation analysis between GW events and AGN catalogs by \citet{Veronesi2023}. The large uncertainties in our inference reflect both the limited number of detections and degeneracies between dynamical channels, underscoring that more stringent constraints will require the larger and more precise samples expected from third-generation detectors.


\subsection{Tracing SMBH growth across cosmic time}
The efficiency of hierarchical mergers in AGN disks depends sensitively on the properties of the host galaxy, particularly on the mass of its SMBH and the length of its cycles of active accretion. 
As a consequence, the cosmic evolution of the SMBH mass function directly impacts the redshift dependence of the AGN-channel BBH merger rate density. We explore two scenarios for SMBH growth, based on \citet{Trinca2022}: Eddington-limited accretion and merger-driven, episodically super-Eddington episodes.

By combining the redshift-dependent SMBH mass function from these models with the cosmic evolution of the AGN number density $n_\mathrm{AGN}(z)$ derived from cosmological 
simulations \citep{Nelson2019}, we predict the BBH merger rate density $\mathcal{R}(z)$ for the AGN channel. In the merger-driven scenario, the early build-up of high-mass SMBHs leads to an 
excess of AGN-assisted mergers at $z \gtrsim 3$ compared to the Eddington-limited case, whereas at lower redshift the two models behave in a similar way. 
Third-generation detectors such as the Einstein Telescope and Cosmic Explorer will extend the reach of BBH observations to high redshift, enabling direct measurements of the AGN-channel merger rate density in the epoch of SMBH assembly. A detection (or absence) of a high-$z$ excess in $\mathcal{R}(z)$ will thus provide a novel observational probe of SMBH growth, potentially distinguishing between slow Eddington-limited accretion and rapid, merger-driven growth across cosmic time.



\section*{Acknowledgments}
The author thanks her collaborators M. Mapelli, A.A. Trani, Y. Seif, D. Wylezalek, A. Trinca, S. Torniamenti, and M. Dall'Amico. She gratefully acknowledges financial support from the MODEST-25 organizers.
She also acknowledges support from the European Research Council for the ERC Consolidator grant DEMOBLACK, under contract no. 770017; from the German Excellence Strategy via the Heidelberg Cluster of Excellence (EXC 2181 - 390900948) STRUCTURES; from the state of Baden-W\"urttemberg through bwHPC and the German Research Foundation (DFG) through grants INST 35/1597-1 FUGG and INST 35/1503-1 FUGG. 
The public version of \textsc{fastcluster} is available via gitlab following \href{https://gitlab.com/micmap/fastcluster_open}{this link}.

\end{document}